\documentclass[12pt,a4paper,prb]{revtex4}

\usepackage[final]{graphicx}
\usepackage{t1enc}
\usepackage{times}

\newcommand{\beq}{\begin{equation}}
\newcommand{\eeq}{\end{equation}}
\newcommand{\beqa}{\begin{eqnarray}}
\newcommand{\eeqa}{\end{eqnarray}}
\newcommand{\veck}{\vec{k}}
\newcommand{\vecr}{\vec{r}}

\begin{document}

\title{\bf Fickian crossover and lengthscales from two point functions in
supercooled liquids}

\author{Daniel A. Stariolo}
\email{stariolo@if.ufrgs.br}
\homepage{http://www.if.ufrgs.br/~stariolo}
\affiliation{Departamento de Física, Universidade Federal do Rio Grande do Sul,
CP 15051, 91501-970 Porto Alegre, Brazil}
\author{Gabriel Fabricius}
\email{fabriciu@fisica.unlp.edu.ar}
\affiliation{Departamento de Física, Universidad Nacional de La Plata, cc 67, 1900
La Plata, Argentina.}
\altaffiliation{Research Associate of the Abdus Salam
International Center for Theoretical Physics, Strada Costiera 11, Trieste,
Italy}

\date{\today}

\begin{abstract} 
Particle motion of a Lennard-Jones supercooled liquid near the glass transition is studied by
molecular dynamics simulations. We analyze the wave vector dependence of relaxation times in
the incoherent self scattering function and show that at least three different regimes can be
identified and its scaling properties determined. The transition from one regime to another
happens at characteristic length scales. The lengthscale associated with the onset of Fickian
diffusion corresponds to the maximum size of heterogeneities in the system, and the characterisitic timescale
is several times larger than the alpha relaxation time. A second crossover lengthscale is observed,
 which corresponds to the typical time and length of heterogeneities, in agreement with results
from four point functions. The different regimes can be traced back in the behavior of the van
Hove distribution of displacements, which shows a characteristic exponential regime in the
heterogeneous region before the crossover to gaussian diffusion and should be observable in
experiments.
 Our results show that it is possible to obtain characteristic length scales
of heterogeneities through the computation of two point functions at different times.

\end{abstract}


\maketitle

\section{Introduction}
The basic mechanisms of diffusion in supercooled liquids and glasses are still poorly understood.
Despite a considerable amount of work in the last years, different approaches to the problem do not
always agree and raise doubts as to the ultimate relevance of the approaches themselves. 
In the landscape 
approach~\cite{De96,DeSt01,Ca01,GrCaGiPa02,BuHe00,DoHe03-3,AnRuSaSc03,AnDiRuScSc00,Wa03} the
basic hypothesis is that, at low temperatures, the dynamics of a supercooled liquid or a glass is
ruled by the complex topography of the potential energy surface, in particular by the multiplicity
and complexity of local minima and saddles. The glass transition in this context can be viewed as
the manifestation of a particular topological transition in the landscape, at which the mechanism
of diffusion by escape through unstable directions ceases to be efficient and hopping through barriers
becomes the dominant mechanism to get out of a restricted region in phase space. While this approach has
been considerably successful and offered deep insights onto the nature of the dynamics in the
supercooled regime, its connection with real
space particle dynamics is not direct, and limited information on the elementary dynamical processes
at the particle level has been obtained. At a more coarse grained level, minimal
models with kinetic constraints have been very successful in order to get intuition and insight
on elementary dynamical processes~\cite{RiSo03,GaCh02,BeGa03_1,JuGaCh04,WhBeGa04,BeChGa05}. In this
kind of models the real space dynamics can be analyzed in great detail and relevant mechanisms of
the microscopic dynamics can be uncovered. There is no underlying landscape and this has raised
the question of the relevance and necessity of a landscape based description of the dynamics of
supercooled liquids. Nevertheless, in spite of their success in identifying many
qualitative features of supercooled liquids, the extreme simplification in their own definition 
poses a limit in the description of many complex phenomena, like aging or thermodynamic properties of
glasses.

Mode Coupling Theory~\cite{Go98,GoSj92,Da04} is perhaps the most successful theoretical approach to 
the dynamics of
the supercooled liquid state. It predicts the existence of relevant regimes for relaxation, like 
the beta and alpha relaxations, and make quantitative predictions for temperatures above the glass
transition. Because of the complex interplay between space and time scales in a
supercooled liquid, most studies based on MCT have focused on time scales at fixed length scales,
e.g. the analysis of the alpha relaxation time scale is usually done focusing on the wave vector
corresponding to the peak of the structure factor. Even the emergence of stretched exponential
relaxation, a benchmark of glassy dynamics, is obtained from MCT only through fitting of numerical
solutions of the complex set of equations describing time correlations. The origin of stretched
exponential relaxation in supercooled liquids is still debated. 
A common scenario opposes a possibly
homogeneously stretched relaxation, which basically means that local regions in a supercooled
liquids relax in the same time scale, with an heterogeneous scenario in which different regions
relax with different characteristic times, probably exponentially, giving rise to a global
stretching of the relaxation. 
Only recently the local dynamics has begun to be accessible experimentally,
through the emergence of new sophisticated experimental techniques~\cite{Ed00,KeBl00,WeCrLeScWe00} 
and the concept of {\em dynamical heterogeneities} has gained force~\cite{Si99,Ed00}. 
It has been realized that complex spatio-temporal correlations characterize an heterogeneous system, 
which can be described by four point functions, correlations between two points in space at two 
different times~\cite{DoFRPaGl02,GlNoSc99,Gl00,ToWyBeBiBo04}. These dynamical susceptibilities show a
peak at a time scale which corresponds to the typical heterogeneity in the system and this
time scale grows when temperature is lowered towards the glass transition.  
An associated
growing length-scale suggests a situation similar to a critical phenomenon. This length-scale is
difficult to measure. Experimental results point to small or moderate values of the characteristic
length of heterogeneities near the glass transition~\cite{Ed00,Si99} and computer simulation
results are inconclusive due to the limitations in the times scales reached by numerical experiments,
when compared to real experiments~\cite{Be04,Gl00}.

A clear signature
of dynamical heterogeneity is the observation of two sets of particles with different degree of
mobility, which allows to define ``slow particles'' and ``fast particles'' over particular
time intervals~\cite{KeBl00,WeCrLeScWe00,PuFuCa04,ReRaGe05,FlSz05}. 
These two sets show up, for example, in a double peak structure of the
distributions of particle displacements, and several indicators have been defined in order to
locate the time at which this separation is maximal. This time scale is another way of defining 
the typical lifetime
of heterogeneities. At very long time scales the supercooled liquid recovers the characteristics
of homogeneous Brownian motion and the distribution of displacements becomes Gaussian. Consequently
indicators of non-Gaussian behavior serve to characterize dynamical heterogeneity~\cite{FlSz05,
SzFl05}.

Recently~\cite{BeChGa05,PaGaCh04}, by performing a detailed study of the interplay between time 
and length scales, it was
realized that it is possible to obtain characteristic length-scales of the process of diffusion in 
supercooled liquids by analyzing two point functions, namely the self part of the van Hove 
distribution function $G_s(\vec{r},t)$ and its Fourier transform, the self incoherent scattering 
function $F_s(\vec{k},t)$, the two most common functions used to study dynamical behavior in 
liquids~\cite{HaMc}. Analyzing the behavior of kinetically constrained lattice models, it was 
realized that
it is possible to define a length-scale corresponding to the onset of Fickian diffusion in a
supercooled liquid. Above this length the system behaves as a simple fluid and
below it persistence dominates and nearly frozen regions are observed up to times corresponding
to the alpha relaxation time, precluding the glass transition. This interesting observation, 
that stretching is dominated by persistence events up to a time in which Fickian diffusion sets
in, can only be a rough approximation to the true behavior of a strongly correlated supercooled
liquid, as we will see in the following. 

In this paper we show results of molecular dynamics simulations of a Lennard-Jones binary mixture 
(LJBM),
focusing on the behavior of the van Hove and self scattering functions through a wide spectrum
of time and length scales~\cite{KoAn95,LiCh01}. Our main result is that characteristic length
scales can be unequivocally obtained through the analysis of the wave vector dependence of
relaxation times in two point time dependent correlation functions. We have been able to 
characterize at least three regimes: ballistic, heterogeneous and Fickian. 
In particular, the scale characteristic of the onset of Fickian diffusion is shown to correspond
to the maximum size of heterogeneities. The other lengthscale corresponds to the typical
heterogeneous behaviour and the corresponding time is the typical relaxation time scale. The
heterogeneous regime is characterized by stretched behavior of the scattering function and we
have observed a corresponding exponential regime in the van Hove distribution function of
displacements.

Our results show that typical
lengthscales of heterogeneities can be obtained through analysis of two point functions, like
the incoherent scattering function, which are easily obtained experimentally.

In section II we introduce the model, the simulation details and show the appearance of the 
different regimes in the inherent structures version of the van Hove distribution function.
In section III we present our main results on the wave vector dependence of relaxation times and
determination of characteristic time and lengthscales from analysis of the self incoherent 
scattering function.
In section IV we make a comparison with known experimental results. 
In section V we present some conclusions.

\section{The Lennard-Jones binary mixture and basic observables}
We performed molecular dynamics simulations on a well known Lennard-Jones binary
mixture (LJBM) with 80 \% particles of type $A$ and  20 \%  particles of type $B$ with
$\epsilon_{AA} = 1.0$,
$\epsilon_{BB} = 0.5$,
$\epsilon_{AB} = 1.5$,
$\sigma_{AA}   = 1.0$,
$\sigma_{BB}   = 0.88$,
$\sigma_{AB}   = 0.8$
at a density 1.204. Most results correspond to systems with N=1000 particles. Some
results with 130 and 10000 particles will be shown for discussing fine size effects.
We used a cutoff radius $R_c=1.8$ for the potential and
obtained $T_{MCT}\approx 0.46$ from extrapolation of diffusivity data. The simulations
shown here were done at a single working temperature $T=0.525=1.14\,T_{MCT}$, at which the system was 
equilibrated. 
As our main aim is to analyze the diffusion dynamics of particles
we map the instantaneous configurations to the nearby local minima, called inherent
structures~\cite{DeSt01,FaSt02}. 
Periodically along a trajectory we take a configuration and let it relax
with a conjugate gradient algorithm to the nearest local minimum. In this way we get
a map of the true trajectory to a trajectory between local minima. 
As the temperature is lowered
towards the glass transition, trapping in the basins of these inherent structures becomes
important and rule the dynamics~\cite{GrCaGiPa02,AnDiRuScSc00,DeReBo02}. Nevertheless
throughout the exposition we make comparisons with the corresponding results from the
real (instantaneous) molecular dynamics trajectories. 

\begin{figure}[ht]
\includegraphics[width=13cm,height=15cm,angle=270]{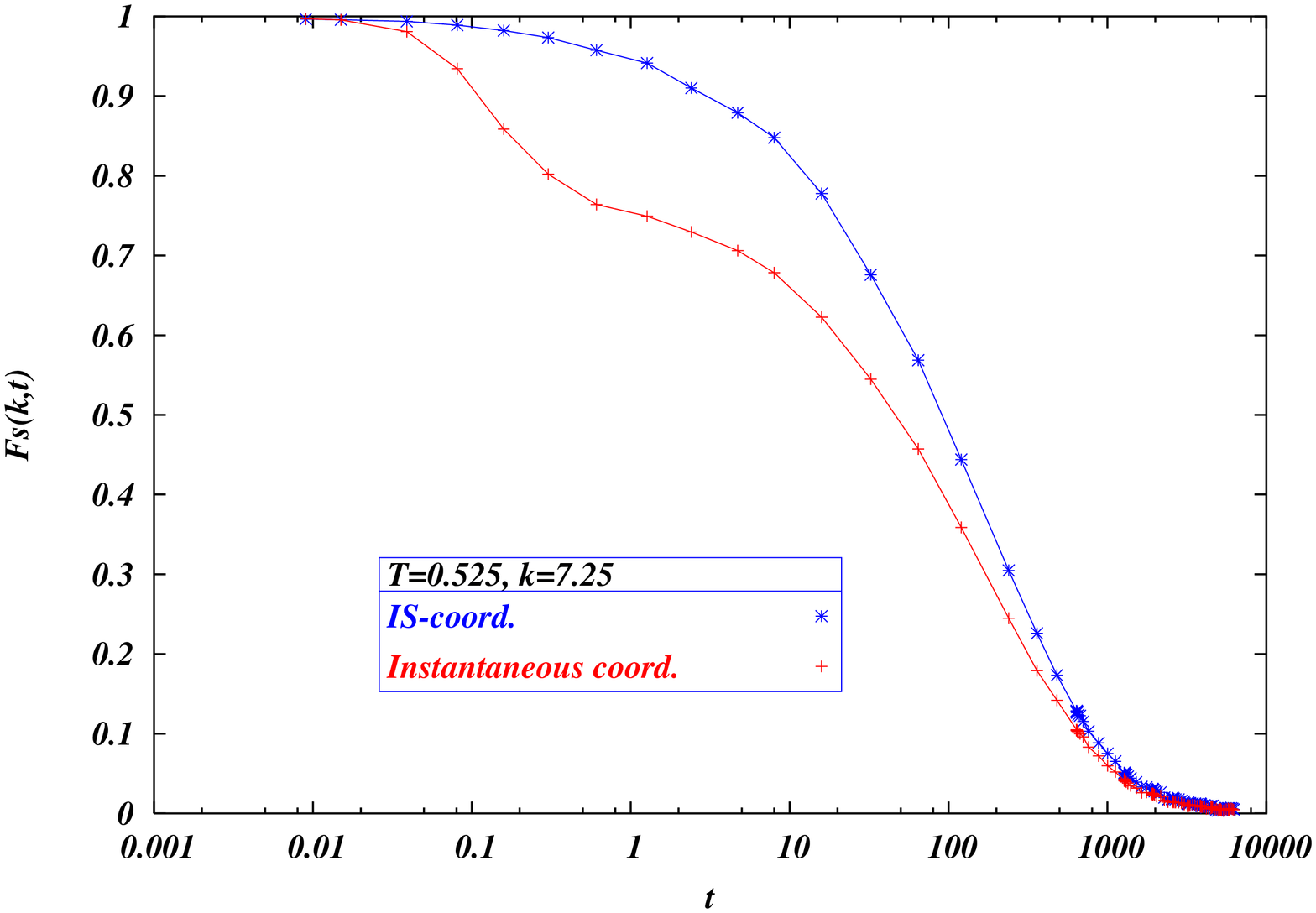}
\label{fig.fsq_T.525}
\end{figure}

In figure \ref{fig.fsq_T.525} we compare the behavior of the {\em self incoherent
scattering function}:
\beq
F_s(\veck,t) = \frac{1}{N}\sum_{i=1}^N \exp{[i\,\veck \cdot(\vecr_i(t)-\vecr_i(0))]}
\eeq
calculated from the real instantaneous configurations, with the same function calculated from
the corresponding inherent structures. From its definition this function is wave vector
dependent. In this figure the wave vector corresponds to the maximum of the structure
factor $|k|=7.25$. The main difference between the two curves is the suppression of the relaxation to the
plateau in the IS dynamics. This basically means that vibrations of the particles and rattling
inside cages is strongly suppressed in the IS dynamics and we are left only with structural
displacements. This is good for our purposes of looking at diffusion dynamics at low temperatures.
From the instantaneous curve we extract as usually the $\alpha$-relaxation time scale which is
approximately $\tau \simeq 120$ for our system.

\begin{figure}[ht]
\includegraphics[width=13cm,height=15cm,angle=270]{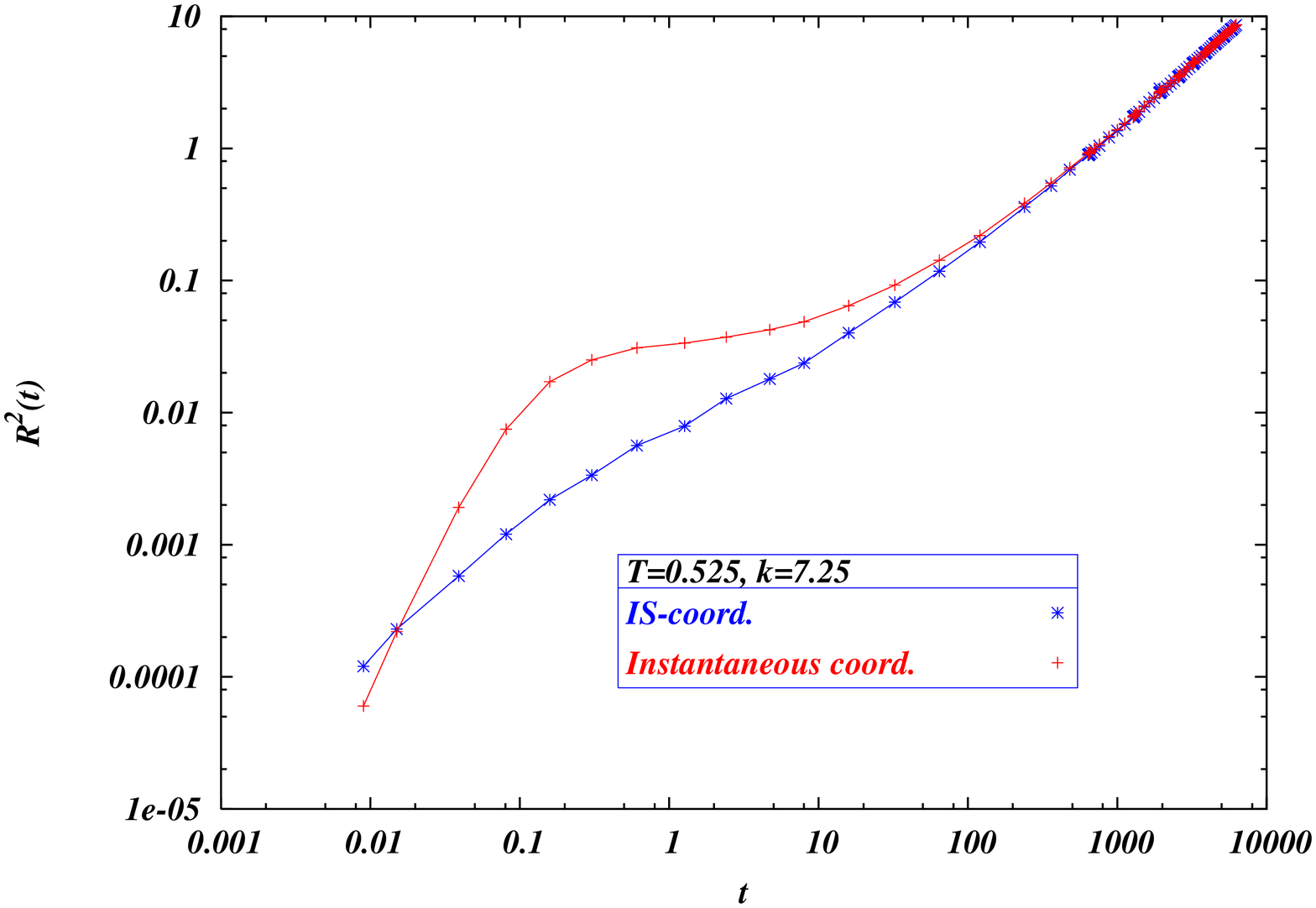}
\label{fig.msd_T.525}
\end{figure}

The same behavior is observed in the mean squared displacement (MSD) of the particles 
$R^2(t)=\frac{1}{N}\sum_{i=1}^{N}|\vecr_i(t)-\vecr_i(0)|^2$ shown in Fig. \ref{fig.msd_T.525}. 
Note that
the MSD from IS proceeds without the arrest at intermediate times. The two curves merge 
approximately around the $\alpha$-relaxation time, from where the system gradually enters a normal
diffusive dynamics.

\begin{figure}[ht]
\includegraphics[width=13cm,height=15cm,angle=270]{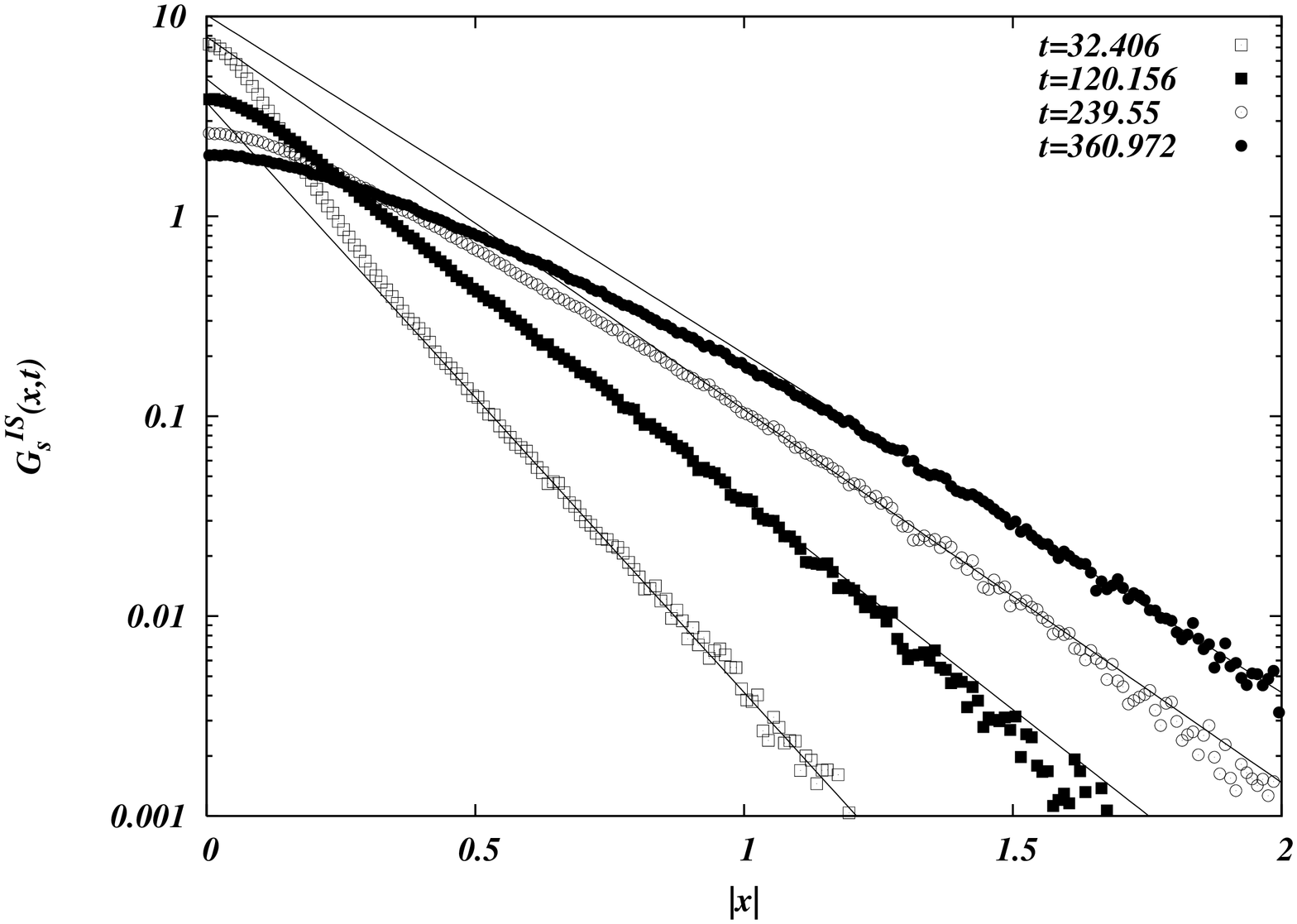}
\label{fig.gs_exp}
\end{figure}

\begin{figure}[ht]
\includegraphics[width=13cm,height=15cm,angle=270]{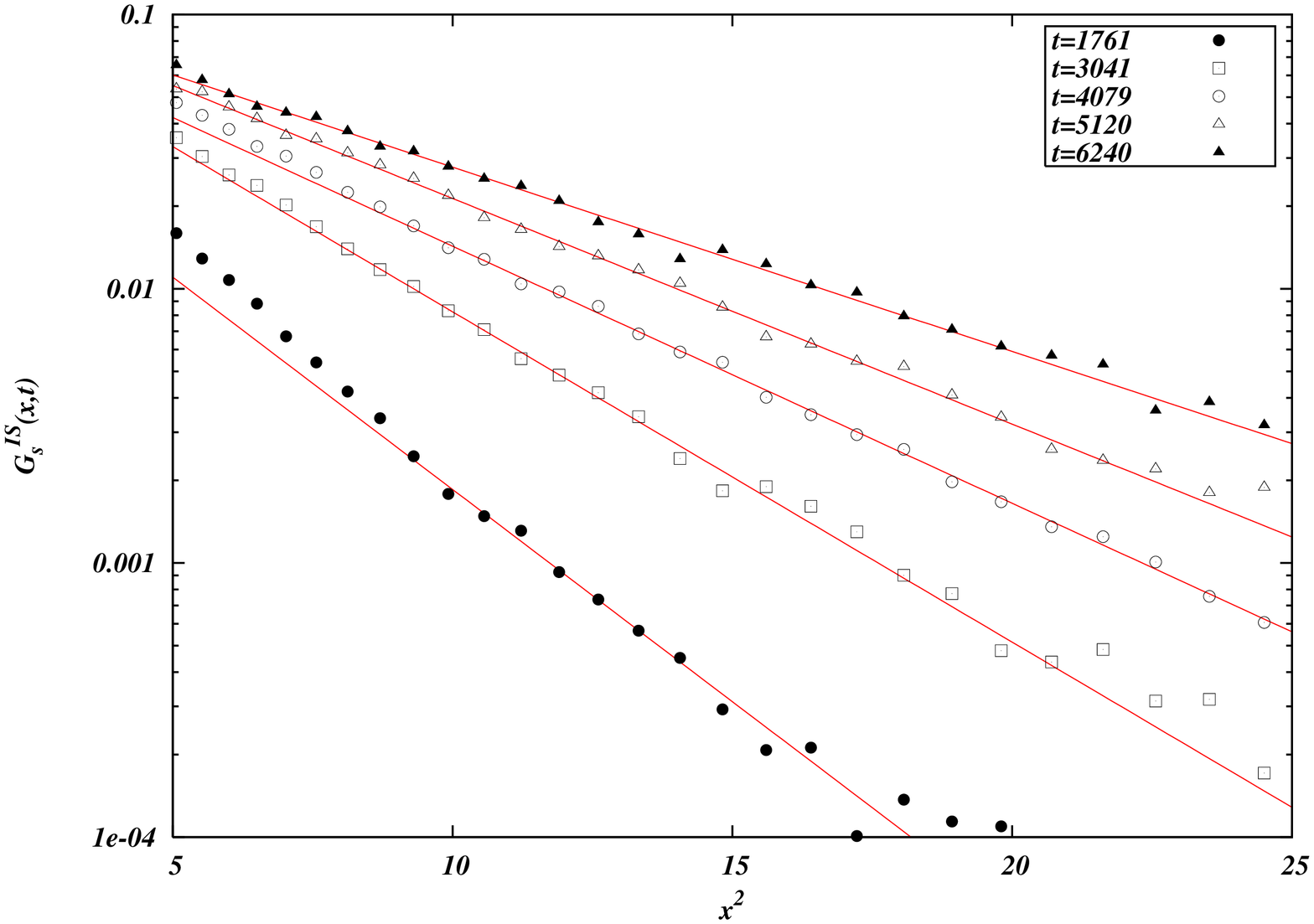}
\label{fig.gs_gauss}
\end{figure}

The basic quantity to characterize the displacements of particles with time is the van
Hove distribution function~\cite{HaMc,KoAn95}. This function has been extensively analyzed in
the context of dynamical heterogeneities in supercooled liquids and colloidal systems
~\cite{KeBl00,WeCrLeScWe00,PuFuCa04,ReRaGe05,FlSz05}. We consider the self part of the function 
and sum the contributions to the displacements in the three coordinate directions obtaining
an effective one dimensional quantity defined by:
\begin{eqnarray}
G_s^{IS}(x,t) & = & \frac{1}{N}\left< \sum_{i=1}^N \delta(x-|x_i^{IS}(t)-x_i^{IS}(0)|) \right> \\
     & \stackrel{t,x\to \infty}{\rightarrow} & \frac{1}{(4\pi Dt)^{1/2}}\ exp\left(
-\frac{x^2}{4Dt} \right) \nonumber \\
\label{vanhove}
\end{eqnarray}
where the superscript {\em IS} means that we are considering inherent structures coordinates.
The long time, large displacement limit corresponds to homogeneous Fickian
diffusion. 
We were able to distinguish at least two regimes. In figure \ref{fig.gs_exp} it is shown that 
an exponential decay fits correctly the data for intermediate time scales and distances $0.5 \leq x \leq 2$.
We will see later when analyzing the wave vector dependence of relaxation times, that this
regime of times and distances corresponds to the heterogeneous or stretched dynamical regime.
An exponential decay in the distribution of displacements was also observed by Vogel et al.~\cite{VoDoHeGl04} 
and Schroeder et al.~\cite{ScSaDyGl00} when analyzing the displacements of particles between two consecutive
inherent structures.
%
In figure \ref{fig.gs_gauss} we show the asymptotic Gaussian regime of the van Hove function.
Note that it is necessary to wait for times several orders longer than the alpha scale in
order that almost all particles enter a Fickian regime. This fact was recently observed
and analyzed in \cite{SzFl05}. The crossover length and time scales between the Fickian
and exponential regimes will be analyzed in grater detail in the next section.

Recently a frozen regime was also
observed in spin models of glasses with kinetic constraints \cite{BeChGa05}. In this regime
a fraction of the system remains frozen up to times which increase rapidly with decreasing
temperature, as described by a persistence function. In the van Hove distribution the frozen
component is reflected by a delta peak at the origin. In molecular systems a strictly frozen
component cannot be seen due to vibrations. But from the inherent structures dynamics we obtained 
a delta peak for our smallest sample of N=130 particles. 
In this case we see that the whole system remains frozen
for some samples up to times of the order of the typical relaxation time. But once a
particle moves, all the others also move, although the displacements can be very small.
As the size of the system grows, completely frozen samples became rare. For the 10000 particles 
system we have not seen the ocurrence of a frozen sample nor of a frozen region within the
samples, i.e. it is not possible to observe strictly frozen groups of particles. 
As the system size grows, the minimum possible displacement of the particles shifts
continuously towards smaller distances. At most we must expect to see groups of particles
with distinctive mobility, i.e. slower and faster particles.
%

\section{The self scattering function and characteristic time-length scales of heterogeneities}

Assuming isotropy of space we computed the following one dimensional self incoherent
scattering function:
\beq
F_s^{IS}(k,t) = \frac{1}{N}\sum_{i=1}^N \exp{[i\,k\,(x_i^{IS}(t)-x_i^{IS}(0))]}
\eeq
in which the superscript {\em IS} means that the function is calculated from inherent
structures configurations.

\begin{figure}[ht]
\includegraphics[width=13cm,height=15cm,angle=270]{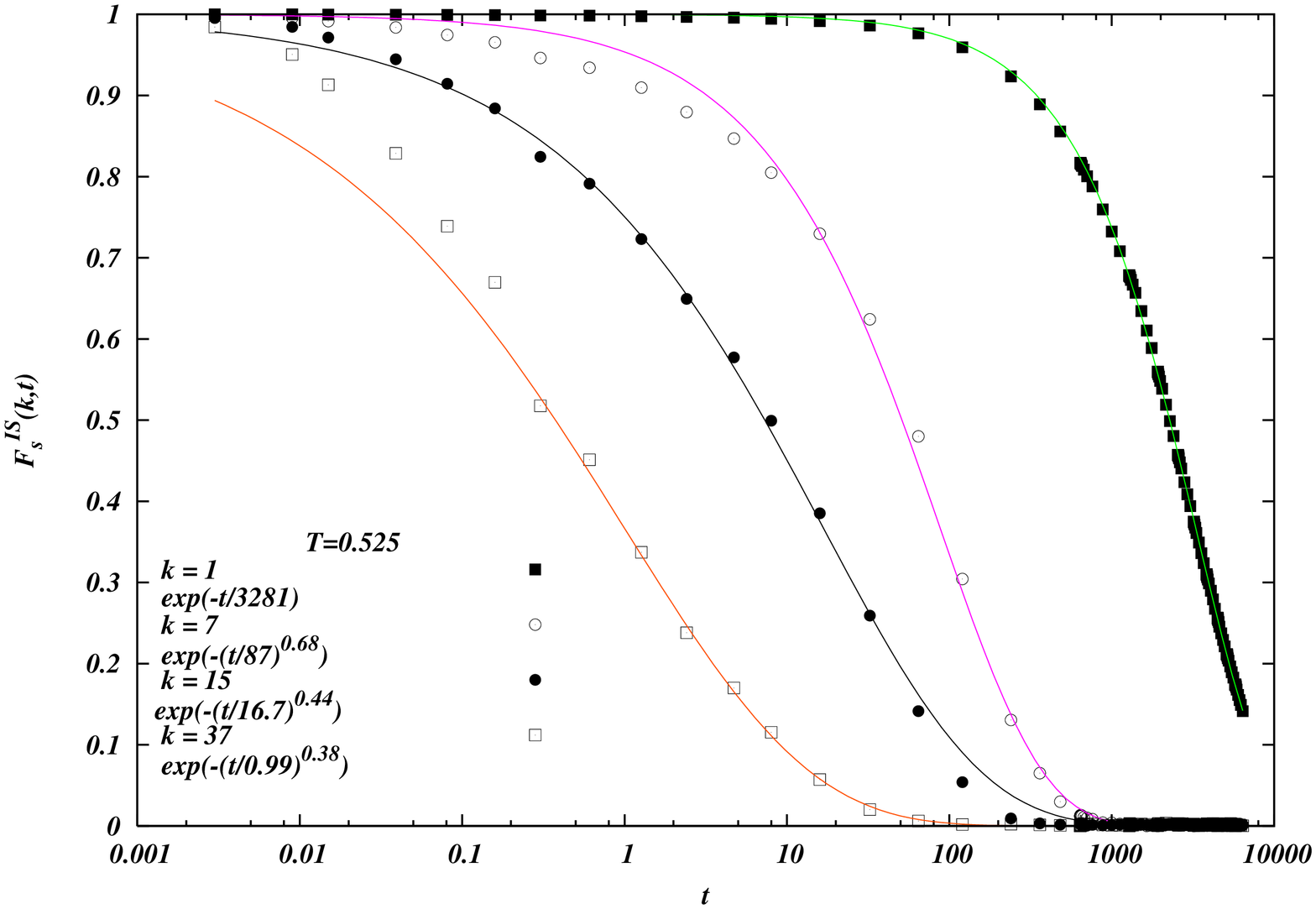}
\label{fig.Fsq.IS_T.525_stretched}
\end{figure}
\begin{figure}[ht]
\includegraphics[width=13cm,height=15cm,angle=270]{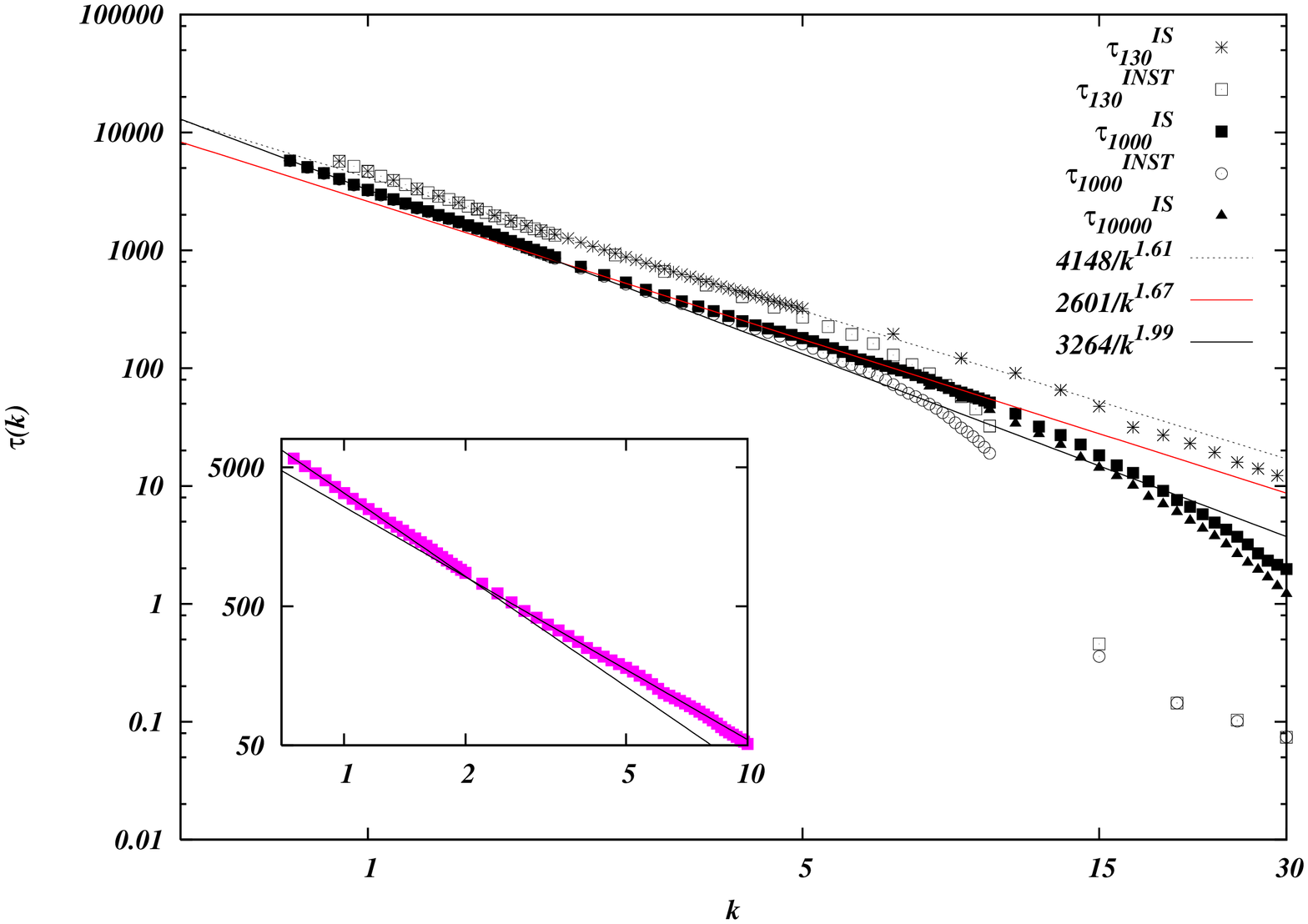}
\label{fig.tau_k}
\end{figure}

In figure \ref{fig.Fsq.IS_T.525_stretched}
we show four selected curves corresponding to different wave vectors in order to illustrate
the different regimes of relaxation. For
small $k$ the relaxation is exponential corresponding to Fickian diffusion of these
modes. For wave vectors around the peak of the structure factor ($k=7$ is shown in the figure)
or larger, the long time decay can be well fitted by a stretched exponential with a $k$ dependent
exponent which will be analyzed below.  The regime of large $k$'s correponds to the system being around
the basin of a single inherent structure, and the decay can be fitted with two stretched
exponentials, one for the short time and another for the long time regimes of the
relaxation. This analysis was performed and discussed in detail in \cite{FaSt04,StArFa04}.
In figure \ref{fig.tau_k} we show the main result of this work,
the dependence of the relaxation time $\tau(k)$ on wave vector. Relaxation times for
each $k$ were obtained as usual, as the time at which the self scattering function decays
to {\em 1/e} of its initial value. In this plot we compare the results from inherent structures (IS)
and real coordinates for three system sizes: $N=130$, $N=1000$ and $N=10000$. We see that the
data for the two larger systems almost coincide, so one can be confident that already for $N=1000$
there are no finite size effects in the scaling regimes. This
figure is very interesting~\footnote{this plot is shown in \cite{BeChGa05} for the
FA kinetic model.}, showing different scaling regimes between time and length scales. 
For each system size and for the smallest wave vectors, the results from IS and real coordinates coincide, 
while they differ strongly at large wave vectors, as expected. 

For wave vectors $k \leq 2$ and times $\tau \geq 1000$, a Gaussian scaling $\tau \simeq 1/k^2$ is observed:
 this is the Fickian diffusion regime (see the inset in figure \ref{fig.tau_k}). 
At $k \approx 2$ a crossover to another scaling form is observed: the
system enters an anomalous diffusion regime with scaling $\tau \simeq 1/k^{1.6}$ . This regime
extends between $2 \leq k \leq 10$ and times between $50 \leq \tau \leq 1000$. This is the
regime in which the dynamics is heterogeneous and the relaxation of correlation functions is
stretched. Comparing the length and time windows in this regime we realize that they correspond
to the exponential decay of the van Hove distribution (see figure \ref{fig.gs_exp}).

For $k \geq 10$ the relaxation times decay rapidly in a narrow interval of wave vectors. Note that
this effect of rapid decay is much more pronounced in the data corresponding to the real coordinates
than in the inherent structures data. This is due to the cage effect, in which the diffusion is
temporarily halted, seen in the figure as a narrow interval in $k$ values. For scales larger than 
$k \geq 20$ (data not shown), a linear scaling works well: the diffusion is ballistic. While the
ballistic regime is nicely observed in the real dynamics, the corresponding curves for the
inherent structures dynamics show instead a slow decay of relaxation times with $k$, because the fast
motion of the particles is filtered out. The second crossover, between the subdiffusion and the
faster decay happens at a time of the order of the alpha relaxation time. This time corresponds to
the peak in the four point susceptibility and the corresponding length scale to the typical size of
heterogeneities.

The crossover between Fickian and anomalous diffusion
at nearly $k_F \simeq 2$ allows to define a characteristic length scale $l_F \simeq 3$ as the
scale for the onset of Fickian diffusion at the temperature considered. Below this scale cooperativity 
rules the motion of individual particles and the dynamics
is heterogeneous. This length scale corresponds to the maximum size of heterogeneities in the system.
The departure from anomalous diffusion at nearly $k_H \simeq 10$ allows to define a second
length scale $l_H \simeq 0.6$, as the typical size of heterogenous regions. Note that this size
is very small at this temperature. It would be extremely interesting to repeat this analysis
for different temperatures approaching and crossing the mode coupling transition temperature.

\begin{figure}[ht]
\includegraphics[width=13cm,height=15cm,angle=270]{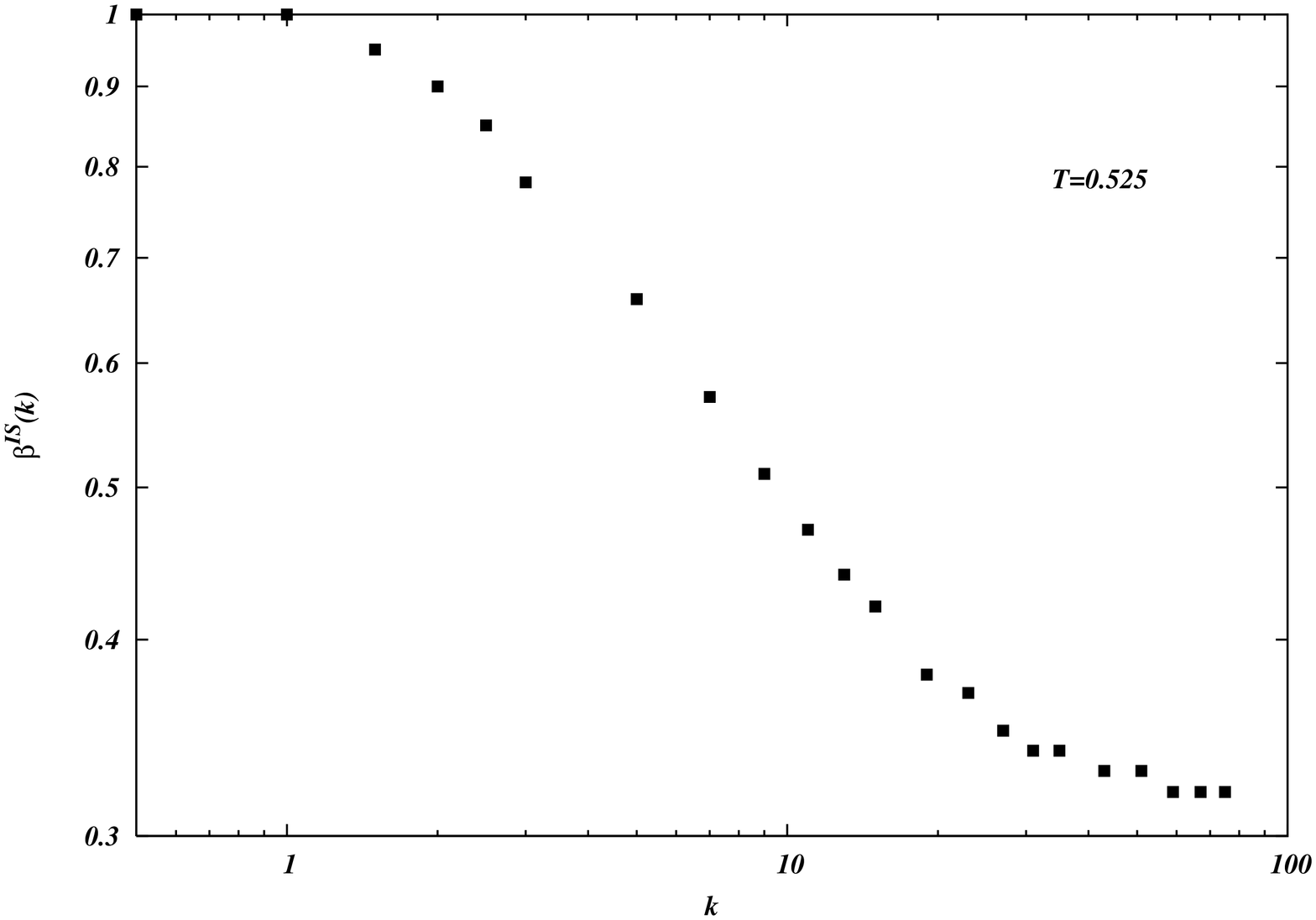}
\label{fig.beta_k}
\end{figure}

In figure \ref{fig.beta_k} we show the dependence of the stretching exponent
$\beta^{IS}(k)$ on wave vector. It decreases monotonously from exponential ($\beta=1$, Fickian) behavior
at small $k$ to an apparent saturation around $\beta \simeq 0.3$ at large $k$. Due to the complex $t$ and
$k$  dependence of the stretching exponent, 
we were not able to collapse the data of the self scattering function for different 
wave vectors onto a single master curve. 



\section{Experimental evidence} \label{experimental}

There is a relatively large literature reporting results on heterogeneous dynamics in glasses
and the search for characteristic time and length scales~\cite{Ed00,Si99}. 
Nevertheless there are, to our knowledge, only a few papers reporting detailed measurements
of particle displacements in supercooled liquids and glasses~\cite{WeCrLeScWe00,KeBl00,BeBiBoCiEl05}.
%

\begin{figure}[ht]
\includegraphics[width=13cm,height=15cm,angle=270]{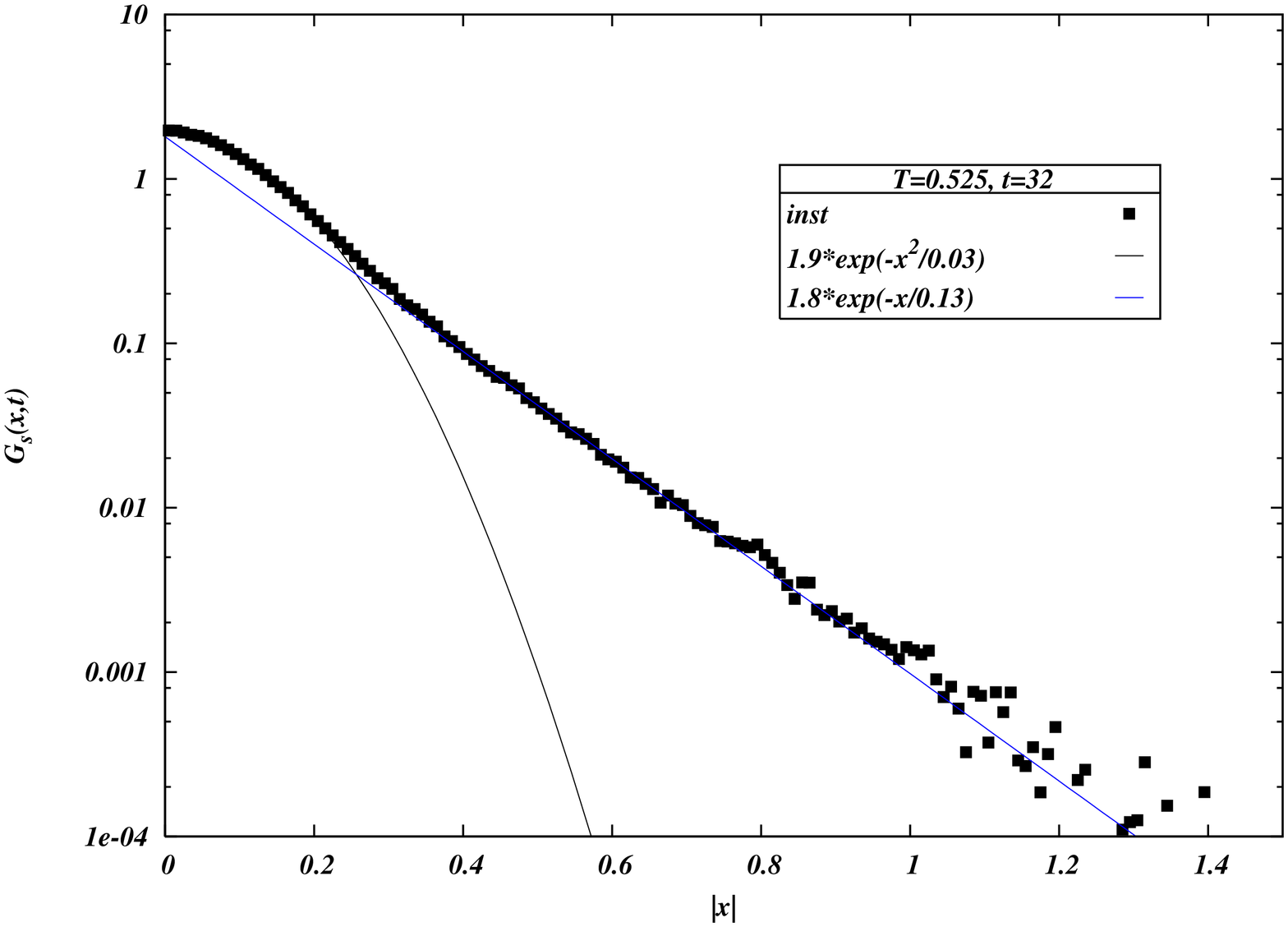}
\label{fig.gs_inst_bb}
\end{figure}

In figure \ref{fig.gs_inst_bb} we show that in the real dynamics, for a time less than the alpha 
relaxation time, the van Hove function can be fitted by a gaussian contribution from vibrations at small
distances plus an exponential contribution at large distances. This figure can be compared,
e.g. with figure 3 of \cite{WeCrLeScWe00}, where experimental data from colloidal supercooled
fluids and glasses was
fitted with an stretched exponential with exponent $\simeq 0.8$, although 
an exponential fit can probably work also in that case.

In another, similar experiment, Kegel et al~\cite{KeBl00}, using time-resolved fluorescence 
confocal scanning microscopy,
also measured the self part of the van Hove function. They classified the particles in two subsets:
one fast and one slow, and fitted the data for both subsets with two Gaussians. The fits are
very good, as recently observed for the same time scale in simulations of the LJBM~\cite{ShDeSt05}.
The difference in the behavior of the distribution function between both experimental results probably
reflects the different time regimes in which the measurements were done in each case. It 
would be very interesting to have accurate measurements in the whole time span between the beta
relaxation scales up to several times the alpha scale in order to get from that an estimation 
of the characteristic length scales discussed in this work.


\section{Conclusion and perspectives}
The mechanisms of particle diffusion in deeply supercooled liquids are still poorly understood. In
particular, the region of intermediate time and length scales, in which diffusion is anomalous, is
still waiting for a complete theoretical description. In the meantime, new highly precise 
measurements probing local dynamics and extensive computer simulations are giving important
insights on the basic mechanisms that underlie particle motion. By means of molecular 
dynamics simulation on a Lennard-Jones supercooled liquid we showed that the
van Hove distribution and self scattering functions still bring us new and rich
 information. They show at least three 
well defined regimes on different time and length scales. 
On very short times and lengths the motion of particles is ballistic. Then heterogeneities develop
and at times of the order of the relaxation time a well defined crossover length $l_H$ can be obtained
from the $k$ dependence of the relaxation times. This length corresponds to the typical size of
spatial heterogeneities. In this region  the diffusion is anomalous and time correlations decay in
a stretched exponential way. Also in this space-time regime the van Hove distribution shows a well 
defined exponential decay. Up to
our knowledge this exponential decay has not been observed nor analyzed in other models, like the much 
studied kinetically constrained lattice models, but is clearly present in experimental results on colloidal
systems. At very long time scales, several orders larger than the alpha relaxation scale, Fickian diffusion
sets in, and the distribution of displacements slowly converges to a Gaussian. The crossover from
heterogeneous to Fickian dynamics is also clearly observed in the $k$ dependence of the self scattering 
function, and a second typical length $l_F$ can be obtained. Scaling forms typical of normal and
anomalous diffusion can be seen for distances larger and smaller than $l_F$, respectively. 
An intersting study on the temperature dependence of the onset time for Fickian diffusion was 
recently done by Szamel et al.~\cite{SzFl05}.
In that work the onset time for Fickian diffusion is defined as the time at which a fixed
deviation from gaussianity was observed in the distribution of the logarithm of particle
displacements $P(log_{10}\delta r,t)$. Its temperature dependence was analyzed and compared with
the corresponding behavior of other characteristic time scales. A particularly interesting
result is the observation that the ratio between the Fickian diffusion onset time $\tau_F$ and
the alpha relaxation time $\tau_{\alpha}$ growths with decreasing temperature, but tends to
saturate near the Mode Coupling transition temperature $T_c$. This may indicate that the
mechanisms behind both time scales are essentially the same as the crossover temperature is
approached.

Although the identification of characteristic dynamical lengths can be naturally introduced through
four point correlation and response functions, we have shown that the relevant information can already be obtained 
from two point functions, like the van Hove distribution and self scattering correlation functions.
To understand the emergence of the different regimes observed in the simulations from a single microscopic model,
 or unified theoretical framework, is still a big challenge.

\begin{acknowledgments}
DAS wish to thank L. Berthier for interesting discussions and comments.
This work was partly supported by Centro Latinoamericano de Física (CLAF), 
CONICET and ANPCyT (Argentina), CNPq and CAPES (Brazil) and ICTP (Italy) through 
grant Net-61 {\em Latinamerican Network on Slow Dynamics in Complex Systems}. 
\end{acknowledgments}


\begin{thebibliography}{43}
\expandafter\ifx\csname natexlab\endcsname\relax\def\natexlab#1{#1}\fi
\expandafter\ifx\csname bibnamefont\endcsname\relax
  \def\bibnamefont#1{#1}\fi
\expandafter\ifx\csname bibfnamefont\endcsname\relax
  \def\bibfnamefont#1{#1}\fi
\expandafter\ifx\csname citenamefont\endcsname\relax
  \def\citenamefont#1{#1}\fi
\expandafter\ifx\csname url\endcsname\relax
  \def\url#1{\texttt{#1}}\fi
\expandafter\ifx\csname urlprefix\endcsname\relax\def\urlprefix{URL }\fi
\providecommand{\bibinfo}[2]{#2}
\providecommand{\eprint}[2][]{\url{#2}}

\bibitem[{\citenamefont{{Debenedetti}}(1996)}]{De96}
\bibinfo{author}{\bibfnamefont{P.}~\bibnamefont{{Debenedetti}}},
  \emph{\bibinfo{title}{Metastable liquids, Concepts and Principles}}
  (\bibinfo{publisher}{Princeton University Press},
  \bibinfo{address}{Princeton}, \bibinfo{year}{1996}).

\bibitem[{\citenamefont{Debenedetti and Stillinger}(2001)}]{DeSt01}
\bibinfo{author}{\bibfnamefont{P.~G.} \bibnamefont{Debenedetti}}
  \bibnamefont{and} \bibinfo{author}{\bibfnamefont{F.~H.}
  \bibnamefont{Stillinger}}, \bibinfo{journal}{Nature}
  \textbf{\bibinfo{volume}{410}}, \bibinfo{pages}{259} (\bibinfo{year}{2001}).

\bibitem[{\citenamefont{Cavagna}(2001)}]{Ca01}
\bibinfo{author}{\bibfnamefont{A.}~\bibnamefont{Cavagna}},
  \bibinfo{journal}{Europhys. Lett.} \textbf{\bibinfo{volume}{53}},
  \bibinfo{pages}{490} (\bibinfo{year}{2001}).

\bibitem[{\citenamefont{Grigera et~al.}(2002)\citenamefont{Grigera, Cavagna,
  Giardina, and Parisi}}]{GrCaGiPa02}
\bibinfo{author}{\bibfnamefont{T.~S.} \bibnamefont{Grigera}},
  \bibinfo{author}{\bibfnamefont{A.}~\bibnamefont{Cavagna}},
  \bibinfo{author}{\bibfnamefont{I.}~\bibnamefont{Giardina}}, \bibnamefont{and}
  \bibinfo{author}{\bibfnamefont{G.}~\bibnamefont{Parisi}},
  \bibinfo{journal}{Phys. Rev. Lett.} \textbf{\bibinfo{volume}{88}},
  \bibinfo{pages}{55502} (\bibinfo{year}{2002}).

\bibitem[{\citenamefont{B\"uchner and Heuer}(2000)}]{BuHe00}
\bibinfo{author}{\bibfnamefont{S.}~\bibnamefont{B\"uchner}} \bibnamefont{and}
  \bibinfo{author}{\bibfnamefont{A.}~\bibnamefont{Heuer}},
  \bibinfo{journal}{Phys. Rev. Lett.} \textbf{\bibinfo{volume}{84}},
  \bibinfo{pages}{2168} (\bibinfo{year}{2000}).

\bibitem[{\citenamefont{Doliwa and Heuer}(2003)}]{DoHe03-3}
\bibinfo{author}{\bibfnamefont{B.}~\bibnamefont{Doliwa}} \bibnamefont{and}
  \bibinfo{author}{\bibfnamefont{A.}~\bibnamefont{Heuer}},
  \bibinfo{journal}{Phys. Rev. Lett.} \textbf{\bibinfo{volume}{91}},
  \bibinfo{pages}{235501} (\bibinfo{year}{2003}).

\bibitem[{\citenamefont{{L. Angelani} et~al.}(2003)\citenamefont{{L. Angelani},
  {G. Ruocco}, {M. Sampoli}, and {F. Sciortino}}}]{AnRuSaSc03}
\bibinfo{author}{\bibnamefont{{L. Angelani}}},
  \bibinfo{author}{\bibnamefont{{G. Ruocco}}},
  \bibinfo{author}{\bibnamefont{{M. Sampoli}}}, \bibnamefont{and}
  \bibinfo{author}{\bibnamefont{{F. Sciortino}}}, \bibinfo{journal}{J. Chem.
  Phys.} \textbf{\bibinfo{volume}{119}}, \bibinfo{pages}{2120}
  (\bibinfo{year}{2003}).

\bibitem[{\citenamefont{Angelani et~al.}(2000)\citenamefont{Angelani, Leonardo,
  Ruocco, Scala, and Sciortino}}]{AnDiRuScSc00}
\bibinfo{author}{\bibfnamefont{L.}~\bibnamefont{Angelani}},
  \bibinfo{author}{\bibfnamefont{R.~D.} \bibnamefont{Leonardo}},
  \bibinfo{author}{\bibfnamefont{G.}~\bibnamefont{Ruocco}},
  \bibinfo{author}{\bibfnamefont{A.}~\bibnamefont{Scala}}, \bibnamefont{and}
  \bibinfo{author}{\bibfnamefont{F.}~\bibnamefont{Sciortino}},
  \bibinfo{journal}{Phys. Rev. Lett.} \textbf{\bibinfo{volume}{85}},
  \bibinfo{pages}{5356} (\bibinfo{year}{2000}).

\bibitem[{\citenamefont{{Wales}}(2003)}]{Wa03}
\bibinfo{author}{\bibfnamefont{D.}~\bibnamefont{{Wales}}},
  \emph{\bibinfo{title}{Energy landscapes}} (\bibinfo{publisher}{Cambridge
  University Press}, \bibinfo{address}{Cambridge}, \bibinfo{year}{2003}).

\bibitem[{\citenamefont{Ritort and Sollich}(2003)}]{RiSo03}
\bibinfo{author}{\bibfnamefont{F.}~\bibnamefont{Ritort}} \bibnamefont{and}
  \bibinfo{author}{\bibfnamefont{P.}~\bibnamefont{Sollich}},
  \bibinfo{journal}{Adv. Phys.} \textbf{\bibinfo{volume}{52}},
  \bibinfo{pages}{219} (\bibinfo{year}{2003}).

\bibitem[{\citenamefont{Garrahan and Chandler}(2002)}]{GaCh02}
\bibinfo{author}{\bibfnamefont{J.~P.} \bibnamefont{Garrahan}} \bibnamefont{and}
  \bibinfo{author}{\bibfnamefont{D.}~\bibnamefont{Chandler}},
  \bibinfo{journal}{Phys. Rev. Lett.} \textbf{\bibinfo{volume}{89}},
  \bibinfo{pages}{035704} (\bibinfo{year}{2002}).

\bibitem[{\citenamefont{Berthier and Garrahan}(2003)}]{BeGa03_1}
\bibinfo{author}{\bibfnamefont{L.}~\bibnamefont{Berthier}} \bibnamefont{and}
  \bibinfo{author}{\bibfnamefont{J.~P.} \bibnamefont{Garrahan}},
  \bibinfo{journal}{Phys. Rev. E} \textbf{\bibinfo{volume}{68}},
  \bibinfo{pages}{041201} (\bibinfo{year}{2003}).

\bibitem[{\citenamefont{Jung et~al.}(2004)\citenamefont{Jung, Garrahan, and
  Chandler}}]{JuGaCh04}
\bibinfo{author}{\bibfnamefont{Y.}~\bibnamefont{Jung}},
  \bibinfo{author}{\bibfnamefont{J.~P.} \bibnamefont{Garrahan}},
  \bibnamefont{and} \bibinfo{author}{\bibfnamefont{D.}~\bibnamefont{Chandler}},
  \bibinfo{journal}{Phys. Rev. E} \textbf{\bibinfo{volume}{69}},
  \bibinfo{pages}{061205} (\bibinfo{year}{2004}).

\bibitem[{\citenamefont{Whitelam et~al.}(2004)\citenamefont{Whitelam, Berthier,
  and Garrahan}}]{WhBeGa04}
\bibinfo{author}{\bibfnamefont{S.}~\bibnamefont{Whitelam}},
  \bibinfo{author}{\bibfnamefont{L.}~\bibnamefont{Berthier}}, \bibnamefont{and}
  \bibinfo{author}{\bibfnamefont{J.~P.} \bibnamefont{Garrahan}},
  \bibinfo{journal}{Phys. Rev. Lett.} \textbf{\bibinfo{volume}{92}},
  \bibinfo{pages}{185705} (\bibinfo{year}{2004}).

\bibitem[{\citenamefont{Berthier et~al.}(2005)\citenamefont{Berthier, Chandler,
  and Garrahan}}]{BeChGa05}
\bibinfo{author}{\bibfnamefont{L.}~\bibnamefont{Berthier}},
  \bibinfo{author}{\bibfnamefont{D.}~\bibnamefont{Chandler}}, \bibnamefont{and}
  \bibinfo{author}{\bibfnamefont{J.~P.} \bibnamefont{Garrahan}},
  \bibinfo{journal}{Europhys. Lett.} \textbf{\bibinfo{volume}{69}},
  \bibinfo{pages}{230} (\bibinfo{year}{2005}).

\bibitem[{\citenamefont{G\"otze}(1998)}]{Go98}
\bibinfo{author}{\bibfnamefont{W.}~\bibnamefont{G\"otze}},
  \bibinfo{journal}{Condensed Matter Physics} \textbf{\bibinfo{volume}{1}},
  \bibinfo{pages}{873} (\bibinfo{year}{1998}).

\bibitem[{\citenamefont{G\"otze and Sj\"ogren}(1992)}]{GoSj92}
\bibinfo{author}{\bibfnamefont{W.}~\bibnamefont{G\"otze}} \bibnamefont{and}
  \bibinfo{author}{\bibfnamefont{L.}~\bibnamefont{Sj\"ogren}},
  \bibinfo{journal}{Rep. Prog. Phys.} \textbf{\bibinfo{volume}{55}},
  \bibinfo{pages}{241} (\bibinfo{year}{1992}).

\bibitem[{\citenamefont{Das}(2004)}]{Da04}
\bibinfo{author}{\bibfnamefont{S.~P.} \bibnamefont{Das}},
  \bibinfo{journal}{Rev. Mod. Phys.} \textbf{\bibinfo{volume}{76}},
  \bibinfo{pages}{785} (\bibinfo{year}{2004}).

\bibitem[{\citenamefont{Ediger}(2000)}]{Ed00}
\bibinfo{author}{\bibfnamefont{M.~D.} \bibnamefont{Ediger}},
  \bibinfo{journal}{Ann. Rev. Phys. Chem.} \textbf{\bibinfo{volume}{51}},
  \bibinfo{pages}{99} (\bibinfo{year}{2000}).

\bibitem[{\citenamefont{Kegel and van Blaaderen}(2000)}]{KeBl00}
\bibinfo{author}{\bibfnamefont{W.~K.} \bibnamefont{Kegel}} \bibnamefont{and}
  \bibinfo{author}{\bibfnamefont{A.}~\bibnamefont{van Blaaderen}},
  \bibinfo{journal}{Science} \textbf{\bibinfo{volume}{287}},
  \bibinfo{pages}{290} (\bibinfo{year}{2000}).

\bibitem[{\citenamefont{Weeks et~al.}(2000)\citenamefont{Weeks, Crocker,
  Levitt, Schofield, and Weitz}}]{WeCrLeScWe00}
\bibinfo{author}{\bibfnamefont{E.~R.} \bibnamefont{Weeks}},
  \bibinfo{author}{\bibfnamefont{J.~C.} \bibnamefont{Crocker}},
  \bibinfo{author}{\bibfnamefont{A.~C.} \bibnamefont{Levitt}},
  \bibinfo{author}{\bibfnamefont{A.}~\bibnamefont{Schofield}},
  \bibnamefont{and} \bibinfo{author}{\bibfnamefont{D.~A.} \bibnamefont{Weitz}},
  \bibinfo{journal}{Science} \textbf{\bibinfo{volume}{287}},
  \bibinfo{pages}{627} (\bibinfo{year}{2000}).

\bibitem[{\citenamefont{{H. Sillescu}}(1999)}]{Si99}
\bibinfo{author}{\bibnamefont{{H. Sillescu}}}, \bibinfo{journal}{J. Non-Cryst.
  Solids} \textbf{\bibinfo{volume}{243}}, \bibinfo{pages}{81}
  (\bibinfo{year}{1999}).

\bibitem[{\citenamefont{Donati et~al.}(2002)\citenamefont{Donati, Franz,
  Parisi, and Glotzer}}]{DoFRPaGl02}
\bibinfo{author}{\bibfnamefont{C.}~\bibnamefont{Donati}},
  \bibinfo{author}{\bibfnamefont{S.}~\bibnamefont{Franz}},
  \bibinfo{author}{\bibfnamefont{G.}~\bibnamefont{Parisi}}, \bibnamefont{and}
  \bibinfo{author}{\bibfnamefont{S.~C.} \bibnamefont{Glotzer}},
  \bibinfo{journal}{J. Non-Cryst. Solids} \textbf{\bibinfo{volume}{307}},
  \bibinfo{pages}{215} (\bibinfo{year}{2002}).

\bibitem[{\citenamefont{Glotzer et~al.}(2000)\citenamefont{Glotzer, Novikov,
  and Schroder}}]{GlNoSc99}
\bibinfo{author}{\bibfnamefont{S.~C.} \bibnamefont{Glotzer}},
  \bibinfo{author}{\bibfnamefont{V.~N.} \bibnamefont{Novikov}},
  \bibnamefont{and} \bibinfo{author}{\bibfnamefont{T.~B.}
  \bibnamefont{Schroder}}, \bibinfo{journal}{J. Chem. Phys.}
  \textbf{\bibinfo{volume}{112}}, \bibinfo{pages}{509} (\bibinfo{year}{2000}).

\bibitem[{\citenamefont{{S. C. Glotzer}}(2000)}]{Gl00}
\bibinfo{author}{\bibnamefont{{S. C. Glotzer}}}, \bibinfo{journal}{J.
  Non-Cryst. Solids} \textbf{\bibinfo{volume}{274}}, \bibinfo{pages}{342}
  (\bibinfo{year}{2000}).

\bibitem[{\citenamefont{Toninelli et~al.}(2005)\citenamefont{Toninelli, Wyart,
  Berthier, Biroli, and Bouchaud}}]{ToWyBeBiBo04}
\bibinfo{author}{\bibfnamefont{C.}~\bibnamefont{Toninelli}},
  \bibinfo{author}{\bibfnamefont{M.}~\bibnamefont{Wyart}},
  \bibinfo{author}{\bibfnamefont{L.}~\bibnamefont{Berthier}},
  \bibinfo{author}{\bibfnamefont{G.}~\bibnamefont{Biroli}}, \bibnamefont{and}
  \bibinfo{author}{\bibfnamefont{J.~P.} \bibnamefont{Bouchaud}},
  \bibinfo{journal}{Phys. Rev. E} \textbf{\bibinfo{volume}{71}},
  \bibinfo{pages}{041505} (\bibinfo{year}{2005}).

\bibitem[{\citenamefont{Berthier}(2004)}]{Be04}
\bibinfo{author}{\bibfnamefont{L.}~\bibnamefont{Berthier}},
  \bibinfo{journal}{Phys. Rev. E} \textbf{\bibinfo{volume}{69}},
  \bibinfo{pages}{020201(R)} (\bibinfo{year}{2004}).

\bibitem[{\citenamefont{Puertas et~al.}(2004)\citenamefont{Puertas, Fuchs, and
  Cates}}]{PuFuCa04}
\bibinfo{author}{\bibfnamefont{A.~M.} \bibnamefont{Puertas}},
  \bibinfo{author}{\bibfnamefont{M.}~\bibnamefont{Fuchs}}, \bibnamefont{and}
  \bibinfo{author}{\bibfnamefont{M.~E.} \bibnamefont{Cates}},
  \bibinfo{journal}{J. Chem. Phys.} \textbf{\bibinfo{volume}{121}},
  \bibinfo{pages}{2813} (\bibinfo{year}{2004}).

\bibitem[{\citenamefont{Reichman et~al.}(2005)\citenamefont{Reichman, Rabani,
  and Geissler}}]{ReRaGe05}
\bibinfo{author}{\bibfnamefont{D.~R.} \bibnamefont{Reichman}},
  \bibinfo{author}{\bibfnamefont{E.}~\bibnamefont{Rabani}}, \bibnamefont{and}
  \bibinfo{author}{\bibfnamefont{P.~L.} \bibnamefont{Geissler}},
  \bibinfo{journal}{J. Phys. Chem. B} \textbf{\bibinfo{volume}{109}},
  \bibinfo{pages}{14654} (\bibinfo{year}{2005}).

\bibitem[{\citenamefont{Flenner and Szamel}(2005)}]{FlSz05}
\bibinfo{author}{\bibfnamefont{E.}~\bibnamefont{Flenner}} \bibnamefont{and}
  \bibinfo{author}{\bibfnamefont{G.}~\bibnamefont{Szamel}},
  \bibinfo{journal}{Phys. Rev. E} \textbf{\bibinfo{volume}{72}},
  \bibinfo{pages}{011205} (\bibinfo{year}{2005}).

\bibitem[{\citenamefont{Szamel and Flenner}(2005)}]{SzFl05}
\bibinfo{author}{\bibfnamefont{G.}~\bibnamefont{Szamel}} \bibnamefont{and}
  \bibinfo{author}{\bibfnamefont{E.}~\bibnamefont{Flenner}},
  \bibinfo{journal}{cond-mat/0508108}  (\bibinfo{year}{2005}).

\bibitem[{\citenamefont{Pan et~al.}(2004)\citenamefont{Pan, Garrahan, and
  Chandler}}]{PaGaCh04}
\bibinfo{author}{\bibfnamefont{A.~C.} \bibnamefont{Pan}},
  \bibinfo{author}{\bibfnamefont{J.~P.} \bibnamefont{Garrahan}},
  \bibnamefont{and} \bibinfo{author}{\bibfnamefont{D.}~\bibnamefont{Chandler}},
  \bibinfo{journal}{cond-mat/0410525}  (\bibinfo{year}{2004}).

\bibitem[{\citenamefont{Hansen and McDonald}(1994)}]{HaMc}
\bibinfo{author}{\bibfnamefont{J.~P.} \bibnamefont{Hansen}} \bibnamefont{and}
  \bibinfo{author}{\bibfnamefont{I.~R.} \bibnamefont{McDonald}},
  \emph{\bibinfo{title}{Theory of {S}imple {L}iquids}}
  (\bibinfo{publisher}{Academic Press}, \bibinfo{year}{1994}),
  \bibinfo{edition}{2nd} ed.

\bibitem[{\citenamefont{Kob and Andersen}(1995)}]{KoAn95}
\bibinfo{author}{\bibfnamefont{W.}~\bibnamefont{Kob}} \bibnamefont{and}
  \bibinfo{author}{\bibfnamefont{H.~C.} \bibnamefont{Andersen}},
  \bibinfo{journal}{Phys. Rev. E} \textbf{\bibinfo{volume}{51}},
  \bibinfo{pages}{4626} (\bibinfo{year}{1995}).

\bibitem[{\citenamefont{Liao and Chen}(2001)}]{LiCh01}
\bibinfo{author}{\bibfnamefont{C.~Y.} \bibnamefont{Liao}} \bibnamefont{and}
  \bibinfo{author}{\bibfnamefont{S.~H.} \bibnamefont{Chen}},
  \bibinfo{journal}{Phys. Rev. E} \textbf{\bibinfo{volume}{64}},
  \bibinfo{pages}{031202} (\bibinfo{year}{2001}).

\bibitem[{\citenamefont{{G. Fabricius} and {D. A. Stariolo}}(2002)}]{FaSt02}
\bibinfo{author}{\bibnamefont{{G. Fabricius}}} \bibnamefont{and}
  \bibinfo{author}{\bibnamefont{{D. A. Stariolo}}}, \bibinfo{journal}{Phys.
  Rev. E} \textbf{\bibinfo{volume}{66}}, \bibinfo{pages}{031501}
  (\bibinfo{year}{2002}).

\bibitem[{\citenamefont{Denny et~al.}(2003)\citenamefont{Denny, Reichman, and
  Bouchaud}}]{DeReBo02}
\bibinfo{author}{\bibfnamefont{R.~A.} \bibnamefont{Denny}},
  \bibinfo{author}{\bibfnamefont{D.~R.} \bibnamefont{Reichman}},
  \bibnamefont{and} \bibinfo{author}{\bibfnamefont{J.-P.}
  \bibnamefont{Bouchaud}}, \bibinfo{journal}{Phys. Rev. Lett.}
  \textbf{\bibinfo{volume}{90}}, \bibinfo{pages}{025503}
  (\bibinfo{year}{2003}).

\bibitem[{\citenamefont{{M Vogel} et~al.}(2004)\citenamefont{{M Vogel}, {B
  Doliwa}, {A Heuer}, and {S Glotzer}}}]{VoDoHeGl04}
\bibinfo{author}{\bibnamefont{{M Vogel}}}, \bibinfo{author}{\bibnamefont{{B
  Doliwa}}}, \bibinfo{author}{\bibnamefont{{A Heuer}}}, \bibnamefont{and}
  \bibinfo{author}{\bibnamefont{{S Glotzer}}}, \bibinfo{journal}{J. Chem.
  Phys.} \textbf{\bibinfo{volume}{120}}, \bibinfo{pages}{4404}
  (\bibinfo{year}{2004}).

\bibitem[{\citenamefont{{T B Schroder} et~al.}(2000)\citenamefont{{T B
  Schroder}, {S. Sastry}, {J. C. Dyre}, and {S C Glotzer}}}]{ScSaDyGl00}
\bibinfo{author}{\bibnamefont{{T B Schroder}}},
  \bibinfo{author}{\bibnamefont{{S. Sastry}}},
  \bibinfo{author}{\bibnamefont{{J. C. Dyre}}}, \bibnamefont{and}
  \bibinfo{author}{\bibnamefont{{S C Glotzer}}}, \bibinfo{journal}{J. Chem.
  Phys.} \textbf{\bibinfo{volume}{112}}, \bibinfo{pages}{9834}
  (\bibinfo{year}{2000}).

\bibitem[{\citenamefont{Fabricius and Stariolo}(2004)}]{FaSt04}
\bibinfo{author}{\bibfnamefont{G.}~\bibnamefont{Fabricius}} \bibnamefont{and}
  \bibinfo{author}{\bibfnamefont{D.~A.} \bibnamefont{Stariolo}},
  \bibinfo{journal}{Physica A} \textbf{\bibinfo{volume}{331}},
  \bibinfo{pages}{90} (\bibinfo{year}{2004}).

\bibitem[{\citenamefont{{D. A. Stariolo} et~al.}(2004)\citenamefont{{D. A.
  Stariolo}, {J. J. Arenzon}, and {G. Fabricius}}}]{StArFa04}
\bibinfo{author}{\bibnamefont{{D. A. Stariolo}}},
  \bibinfo{author}{\bibnamefont{{J. J. Arenzon}}}, \bibnamefont{and}
  \bibinfo{author}{\bibnamefont{{G. Fabricius}}}, \bibinfo{journal}{Physica A}
  \textbf{\bibinfo{volume}{340}}, \bibinfo{pages}{316} (\bibinfo{year}{2004}).

\bibitem[{\citenamefont{{L Berthier} et~al.}(2005)\citenamefont{{L Berthier},
  {G Biroli}, {J-P Bouchaud}, {L Cipelletti}, {D El Masri}, {D L'Hôte}, {F
  Ladieu}, and {M Pierno}}}]{BeBiBoCiEl05}
\bibinfo{author}{\bibnamefont{{L Berthier}}}, \bibinfo{author}{\bibnamefont{{G
  Biroli}}}, \bibinfo{author}{\bibnamefont{{J-P Bouchaud}}},
  \bibinfo{author}{\bibnamefont{{L Cipelletti}}},
  \bibinfo{author}{\bibnamefont{{D El Masri}}},
  \bibinfo{author}{\bibnamefont{{D L'Hôte}}}, \bibinfo{author}{\bibnamefont{{F
  Ladieu}}}, \bibnamefont{and} \bibinfo{author}{\bibnamefont{{M Pierno}}},
  \bibinfo{journal}{Science} \textbf{\bibinfo{volume}{310}},
  \bibinfo{pages}{1797} (\bibinfo{year}{2005}).

\bibitem[{\citenamefont{{M. Scott Shell} et~al.}(2005)\citenamefont{{M. Scott
  Shell}, {P. G. Debenedetti}, and {F. H. Stillinger}}}]{ShDeSt05}
\bibinfo{author}{\bibnamefont{{M. Scott Shell}}},
  \bibinfo{author}{\bibnamefont{{P. G. Debenedetti}}}, \bibnamefont{and}
  \bibinfo{author}{\bibnamefont{{F. H. Stillinger}}} (\bibinfo{year}{2005}),
  \bibinfo{note}{preprint cond-mat/0506608}.

\end{thebibliography}
%
\newpage
{\Large Captions for figures}
\begin{itemize}
\item Figure 1: Self incoherent scattering function for the LJBM at T=0.525 for a wave vector corresponding
to the peak of the structure
factor k=7.25, from instantaneous and inherent structures coordinates.
\item Figure 2: Mean squared displacement at T=0.525 for instantaneous and inherent structures 
coordinates.
\item Figure 3: The van Hove distribution in the exponential regime. Continuous lines are exponential fits.
\item Figure 4: The van Hove distribution in the Fickian regime. 
Continuous lines are gaussian fits.
\item Figure 5: The self incoherent scattering function for four different wave vectors characteristic
of different scaling regimes. In the legend the corresponding fitting
functions (continuous lines).
\item Figure 6: Wave vector dependence of relaxation times for three different sizes, inherent and
instantaneous dynamics. Different scaling regimes indicated with full lines. Inset: zoom of
the Fickian crossover region.
\item Figure 7: Wave vector dependence of the stretching exponent $\beta$.
\item Figure 8: The van Hove distribution from instantaneous coordinates at $t=32$. The data can
be fitted by a Gaussian at small distances plus an exponential decay at larger distances.
\end{itemize}

%
%
\newpage
\vspace{3cm}
FIGURE 1
\newpage
\vspace{3cm}
FIGURE 2
\newpage
\vspace{3cm}
FIGURE 3
\newpage
\vspace{3cm}
FIGURE 4
\newpage
\vspace{3cm}
FIGURE 5
\newpage
\vspace{3cm}
FIGURE 6
\newpage
\vspace{3cm}
FIGURE 7
\newpage
\vspace{3cm}
FIGURE 8

\end{document}